**On the detectability by novices, radiologists, and computer algorithms of smallest increases in local single dot size in random-dot images**

Birgitta Dresp-Langley[1]

John Wandeto[1,2]

[1]ICube UMR 7357 CNRS University of Strasbourg, 4 rue Blaise Pascal, CS 90032 F-67081 Strasbourg Cedex - FRANCE

[2]Dedan Kimathi University of Technology, Nyeri – Mweiga Road, P.O. Box 657-10100, Nyeri - KENYA

**Abstract**

Time-series of images may reveal important information about changes in medical or environmental conditions, depending on context. Visual inspection of images by humans (experts or laymen) may fail in detecting very small differences between images, yet, small but visually undetectable differences may carry important significance. Computer algorithms may help overcome this problem, and the use of computer driven image analysis in medical practice or for the tracking of small but critical changes in natural environments attracts a lot of interest. In many contexts relevant to society, the preprocessing of large sets of image series will soon no longer be the exclusive realm of a few scientists. Here we show that a metric obtained from self-organizing map analysis (SOM) of image contents in time series of images of one and the same object or environment reliably signals potentially critical local changes in images that may not be detectable visually by a layman or even an expert.

**Keywords:** computer; society; medicine; environment; image time series; random-dot images; change detection; self-organizing visual maps; quantization error; human detection

# 1. Introduction

This report here starts out from a scientific perspective couched in the framework of Kohonen's neural network analysis of visual information in terms of self-organizing maps (SOM), also frequently referred to as Kohonen maps. Recent work by Wandeto et al. (2017) has shown that the use of the quantization error (QE), a metric obtained from self-organizing map analysis (SOM) of image contents as a function of time (time series of images of one and the same object or environment), reliably signals potentially critical local changes in images that may not be detectable visually by a layman or even an expert. Moreover, the approach reported on here does not suffer from the well-known limitations of image segmentation. Previous results by Wandeto et al (2017) had shown that the quantization errors (*QE*) increase systematically and reliably with small local increases in lesions in time series of medical (MRI) images. Results were consistent with previous work by others using alternative approaches. Here, the diagnostic potential of the *QE* is assessed in the light of detection experiments where human non-experts and expert radiologists had to detect very small local differences in random-dot images.

In medicine, the annotation of image data is subject to considerable differences between individuals even when they are highly specialized experts such as radiologists [1]. The analysis of medical images assisted by computer techniques therefore represents a highly complex challenge. Radiologists have to assess the progression of patients' conditions on the basis of often hardly detectable, local changes in medical images. These are captured through various imaging techniques, such as magnetic resonance imaging (MRI), computerized tomography (CT), and positron emission tomography (PET), providing the radiologist with visual information about the state or progression of a given condition and helping to determine the course of treatment. Traditional methods for handling medical images involve direct visual inspection, which is by its nature subjective. Image science therefore has proposed methods for reducing subjectivity by introducing automated procedures. This involves various different image processing techniques aimed at identifying specific diagnostic regions, so-called regions of interest, and specific features representing tumors and lesions. For example, to avoid time-consuming voxel-by-voxel comparison for detecting changes between two images, the images can be aligned and displacement fields may be computed for recovering apparent motion by using a non-rigid registration algorithm [2]. This and similar techniques focus on the detection of regions of interest, with tumors or evolving lesions. A computer algorithm compares multiple series of images to produce a map of the changes, and expert knowledge is then applied to that map in a series of post-processing steps in order to generate a set of metrics describing the changes occurring in the images. During this process, domain-specific knowledge needs to be introduced, which attempts to reduce the impact of subjectivity by incorporating generic information an expert might use when annotating medical images manually. This, however, does not completely eliminate subjectivity. Other approaches [3, 4] have proposed a computational framework to enable comparison of MRI volumes based on gray-scale normalization, to determine quantitative tumor growth between successive time intervals. Specific tumor growth indices were computed, such as volume, maximum radius, and spherical radius. This approach also requires the initial manual segmentation of the images, which is a time-consuming task. Semi-automatically segmenting successive images and then aligning them on the basis of hierarchical registration schemes has also been proposed for measuring growth or shrinkage in local image details [5]. All these methods rely on the accuracy of segmentation and require manual annotation for classifying local changes in pathology of up to a few voxels. Other methods [6, 7] which combine input from a medical expert with a computational technique are more specifically aimed at difficult-to-detect brain tumor changes. These methods, again, involve a subjectivity factor which is problematic given the well-known inter-individual differences between experts [1]. In this paper here, with a similar goal in mind, we introduce a new technique based on the functional principles of self-organized mapping [8, 9]. It considers the whole medical image as opposed to an image segment of a specific region of interest. A method of direct analysis of the medical image as a whole has the advantage of not requiring manual benchmarking. The basic idea behind direct image analysis is that there exists an intrinsic relationship between images with medical contents and their clinical measurements, and that this relationship can be exploited directly without additional, and not necessarily reliable, intermediate procedures of analysis. Compared to some of the traditional methods briefly reviewed here above, this approach here has a deep clinical significance because it is directly targeting the final outcome like a human expert would. It thereby bridges the gap between machine learning and medical image analysis, and it will be shown here that a specific output variable of the SOM, the quantization error (*QE*), can be exploited as a diagnostic indicator for the presence of potentially critical local changes in medical image contents. To highlight the full potential of this new method, *QE* outputs from analyses of random-dot images with small changes in local contents will be compared to the average capacity of thirty two non-expert human observers to detect these changes in the same images.

## 2. Materials and Methods

### 2.1 Self-organizing maps

A self-organizing map (SOM) is an unsupervised neural network learning technique [8, 9] that does not need target outputs required in error correction supervised learning, and is used to produce a lower-dimension representation of the input space. Thus, for each input vector, so called competitive learning is carried out to produce a lower-dimension visualization of the input data. SOMs are typically applied as feature classifiers of input data. From an initial randomization of a map, input data is iteratively applied to optimize the map into stable regions. Where the node weights match the input vector, that area of the lattice is selectively optimized to more closely resemble the data for the class the input vector is a member of. From an initial distribution of random weights and over multiple iterations the SOM eventually settles into a map of stable zones. Each region of the map becomes a feature class of the input space. Each zone is effectively a feature classifier, and the graphical output is a type of feature map of the input space.

FIGURE 1

The central idea behind the principles and mathematics of SOM is that every input data item shall be matched to the closest fitting region of the map, called the winner (as denoted by $M_c$ in Fig. 1), and such subsets of regions will be modified for optimal matching of the entire data set [9]. On the other hand, since the spatial neighborhood around the winner in the map is modified at a time, a degree of local and differential ordering of the map occurs to provide a smoothing action. The local ordering actions will gradually be propagated over the entire SOM. The parameters of the SOM models are variable and are adjusted by learning algorithms such that the maps finally approximate or represent the similarity of the input data. While other studies have mainly concentrated on the performance of various SOM on a given dataset, this study here sets out to unveil the behavior of different datasets (here visual random-dot images) in a single SOM. Given a related set of images of a time series and a constant SOM, it should be possible to detect critical changes in these images that bear clinical significance or reflect the evolution of a condition (medical, environmental and other).

### 2.2 The quantization error of the SOM output

The task of finding a suitable subset that describes and represents a larger set of data vectors is called vector quantization [10]. Vector quantization aims at reducing the number of sample vectors or at substituting them with representative centroids. The resulting centroids do not necessarily have to be from the set of samples but can also be an approximation of the vectors assigned to them, for example their average. Vector quantization is closely related to clustering, and SOM performs vector quantization since the sample vectors are mapped to a (smaller) number of prototype vectors, as demonstrated in [11]. The prototype vectors are called the best matching units (BMU) in SOM. As a property of SOM, the quantization error (*QE*) is used to evaluate the quality of SOM. The *QE* belongs to a type of measures that have been used to benchmark a series of SOMs trained from the same dataset. In the present study, the *QE* is used to perform a somewhat opposite measure, used to benchmark a series of datasets with SOM trained with the same parameters. In other words, the same SOM, same map size, feature size, learning rate and neighborhood radius is used to analyze series of image datasets with clinical significance, or random-dot images, as shown later. The *QE* is derived after subjecting an image to a self-organizing map algorithm analysis and by calculating the squared distance (usually, the standard Euclidean distance) between an input data, *x*, and its corresponding centroid, the so-called "best matching unit", or BMU. This gives the average distance between each data vector (*X*) and its BMU and thus measures map resolution:

$$QE = \frac{1}{N}\sum_{i=1}^{N}\|X_i - (\text{BMU}_{(i)})\| \qquad (1)$$

where N is the number of sample vectors x in the image.

This measure completely disregards map topology and alignment, as noted by [11], making it applicable for different kinds and shapes of SOM maps. Besides, the calculation does not rely on any user parameters as seen in (1) above. A 16 by 16 SOM with an initial neighborhood radius of 5 and learning rate of 0.2 was set up for the extraction of data from images. These initial values were obtained after testing several sizes of the SOM to check that the cluster structures were shown with sufficient resolution and statistical accuracy, as in [8]. The learning process was started with vectors picked randomly from the image array as the initial values of the model vectors. The SOM parameters were kept constant.

3. **Small local changes in medical images**

Wandeto et al [14] used the *QE* from SOM to detect small differences in image time series from MRI scans of a patient's knee taken before and after blunt force traumatic injury, which had produced small local, visually barely detectable changes in the MRIs. Figure 2 shows the *QE* values obtained from time series of 20 images per set taken on two consecutive clinical visits at separate dates, before and after blunt force traumatic injury.

FIGURE 2

The *QE*s were submitted to one-way analysis of variance (ANOVA). The difference between image series is statistically significant (*t* (1, 38) = 3,336; p<.01), which directly reflects the clinical significance of the image differences between the first and the second visit, i.e. the effects of blunt force traumatic injury on the MRI image contents (for details, see Wandeto et al [14]).

**4.     Small local changes in random-dot images**

A systematic increase in the quantization error of the SOM output can also be directly linked to the detectability of potentially critical local image contents in visual image discrimination experiments using a classic "same-different" paradigm, as previously shown by Wandeto et al [14, 15]. Visual random-dot images with different percentages of artificially induced and strictly local "lesion" contents (5%, 10 % and 30 %) were paired with original images where no such local "lesion" was added. On each of these images, SOMs were run to determine the quantization error output and to compare its variation with variations in visual change detectability by unexperienced observers. In this experiment, human observers had to judge whether a given image pair was the "same" or "different". Any detection of a difference, called correct positive or "hit", could only be due to detection of the artificially induced local difference ("lesion" content) in one of the images, as all other image parameters (contrast intensity, contrast sign, spatial distribution of contrasts, relative size) were identical in two images of a pair. To determine the subjects' tendency to over-diagnose, pairs of strictly identical images were also presented and the number of false positive detections, or "guesses" recorded. The exposure duration of the image pairs was varied to test whether the processing time affects detectability. Experiments were run with novices and expert radiologists as subjects.

*4.1. Subjects*

32 healthy, young novices, 26 male and 6 female, all volunteers aged between 19 and 34 years of age and 3 expert radiologists, two male and 1 female, participated in the studies. Experiments were conducted in conformity with the Helskinki Declaration relative to experimental investigations on human subjects and had been fully approved by the ethics board of the supervising author's (BDL) host institution (CNRS). All subjects had normal visual acuity and gave written informed consent to participate.

*4.2. Experimental stimuli and procedure*

Computer generated random-dot images of identical size, local contrast (0.7 Michelson contrast) and spatial contrast distribution were created (see Figure 3 for an illustration) using Adobe RGB in Photoshop. In three of these images, one local contrast dot was increased in diameter yielding one image with a 5% local dot size increase, another one with a 10% local dot size increase, and a third one with a 30% local dot size increase, always at exactly the same dot location.

FIGURE 3

Each of these three images was paired with the original "no lesion" image, presented to the left and the right in a pair, in a random order. Images were also paired with their identical images. During the experiment, the subject was seated at a distance of about 75 centimetres from the computer screen in a semi-dark room. The image pairs were presented in a random sequence and each pair was followed by a blank screen presentation of five seconds to avoid visual afterimages, which could have interfered with the task. In one session, the exposure duration for each image pair was five seconds, in another session, the exposure duration was observer controlled; i.e. the subject could look at a pair for as long as he deemed necessary to reach a decision, before pressing a key to get the five-second blank screen before the next pair was displayed. The task instruction was to "decide as swiftly and accurately as possible whether two images in a pair appear to be the *same* or *different*. The number of "same" and "different" judgments in response to a given image pair was recorded and written into an individual excel table, for each subject and session. 16 of the 32 subjects started with the five second exposure duration session followed by the session with the observer controlled exposure duration, the other 16 performed the task sessions in the reversed order to counterbalance possible sequential timing effects.

## 5.    Results

*5.1.    Conditional detection rates of the novices*

The total number of "same" and "different" responses for each type of image pair was divided by the total number of presentations of that pair for a given subject and experimental session. These response frequencies were then multiplied by 100 to produce percentages of correct negatives (*CN*) reflecting the percentage of "same" responses to pairs of the same image, false negatives (*FN*) reflecting the percentage of "same" responses to pairs of different images, false positives (*FP*) also called "guesses", reflecting the percentage of "different" responses to pairs of the same image, and correct positives (*CP*) also called "hits", reflecting the percentage of "different" responses to pairs of different images. The distributions are shown in Tables 1, 2, and 3 as a function of the "lesion" contents, with 5%, 10% and 30% local increase in single dot size, and as a function of the exposure duration of the image pairs. It was confirmed that the position of an image in a pair (*left* or *right*) had no effect on the responses (no positional bias). Average response frequencies for images positioned on *left* and on *right* are shown here.

TABLE 1

TABLE 2

TABLE 3

When comparing between results shown in a) and b) of Tables 1-3, it is made clear that the percentage of false positives (*FP)*, the so-called "guess rate", does not vary much with the exposure duration of the image pairs, whereas the percentage of correct positives (*CP)*, the so-called "hit rate", increases markedly when the exposure duration is *ad libitum* and observer controlled. This indicates that the subjects used constant decision criteria, otherwise the false positive rate (*FP*) would also have varied with the image exposure duration, in the two successive experimental sessions, and that limiting image exposure times negatively affects the correct positives (*CP*) rate. When comparing between Tables 1-3, it is also quite clear that the correct positives rate increases as the "lesion" content in one of the images of a pair increases. In pairs where one of the images has a 5% local dot size increase (Table 1), the rate of correct positives (*CP*) is smaller than the rate of false positives (*FP),* which indicates that the subjects are basically guessing and are unable to detect the local difference in image contents. In pairs where one of the images has a 10% or a 30% local dot size increase, the rate of correct positives (*CP)* is twice (Table 2) to three times (Table 3) the rate of false positives (*FP)*, which shows that the local difference in the image contents is beginning to be detected every now and again, however, well below the level of a psychophysical 75% correct detection threshold. In pairs with observer controlled exposure duration for 30% increase in local dot size, the correct positives rate (*CP*) is the highest at 40%. The 75% correct psychophysical threshold is never attained, which leads to conclude that the changes in the images are not reliably detectable by the human visual system of non-experts.

*5.2. Analysis of variance on the non-expert detection data*

In a next step, the average correct positive (*CP*) rates were submitted to Two-Way ANOVA for the three levels of the "lesion" factor $L_3$ and the two levels of the exposure duration factor $E_2$ to assess the statistical significance of the effects. A statistically significant result was observed for the effect of "lesion" on the average correct positives rates, with $F(2, 23) = 38.04$; $p<.001$, and also a significant effect of "exposure duration", with $F(1, 23) = 8.13$; $p<.05$. The effect sizes in terms of means and standard errors (SEM) are graphically represented in Figure 4.

FIGURE 4

*5. 3. Comparison with QE values from SOM*

To compare the human detection rates with the *QE* values from the SOM analyses run on the random-dot images with 5%, 10% and 30% increase in single local dot size, these *QE* values are shown graphically here in Figure 5 as a function of each image type.

FIGURE 5

*5. 4. Conditional detection rates of the three expert radiologists*

With a 5% local dot size increase (Table 5), there is no detection even by the expert radiologists who have a lot of experience in scanning complex images visually for changes in fine detail. Here, the experts are basically guessing (see their CP rate after subtraction of the FP or "guess" rate). Note that the guess rate (FP rate) of the three experts here is noticeably higher than was that of the novices. Expert radiologists thus seem to have a stronger tendency to produce false positives ("better safe than sorry" strategy).

TABLE 5

With a 10% local dot size increase (Table 6), the experts are still basically guessing, but beginning to detect the difference in the two images when they can control the exposure duration (the CP rate – FP or "guess" rate is no longer zero, but about 25).

TABLE 6

With a 30% local dot size increase (Table 7), the experts in the same way as the novices better detect the difference in local dot size between the two images when they can control exposure duration. The CP rate – FP rate or "guess" rate is 33.3. Given the high false positive (FP), or "guess" rates, of the experts, who tend to respond 'different' when the two images are the same to a greater extent than the novices, we cannot conclude that the experts more reliably detect the 30% increase in local lesion content. Their correct positive rate after subtraction of the false negative rate is still well below the psychophysical threshold of 75% correct.

TABLE 7

## 6. Discussion

In previous research [7], it was reported that an expert wrongly classified all cases with 1% artificial lesion growth, and only achieved an accuracy of 20% for cases with 5% growth. The same expert correctly classified all cases with a 22% growth. In this study here, a new SOM-based technique is introduced for automatically sensing the progression or remission of lesions in medical images. It is demonstrated that the *QE* of the SOM output of consecutive analyses of sets of images taken over a time series increases when impurities/lesions on the organ have increased and vice versa. The human detection data on the random dot images here show quite clearly that minimal growth in local image contents that is not even detected at the psychophysical threshold level of 75% by human non-experts is consistently and reliably captured by the technique introduced in this work. It is useful to recall here that even a 75% correct detection level would still be largely insufficient in a medical context, where the desired detection threshold is 100%. Also, it would make no sense trying to link the percentage of human detection to the percentage of increase in the *QE* measure. The *QE* measure is a numerical indicator of change between zero and infinity and has no upper or lower threshold limit. Small changes in this indicator in response to changes in image contents may reflect a statistically highly significant result with a clear clinical significance, illustrated by the results from SOM analysis on the original medical image series here in this study. The detection rate of the human visual system is of an entirely different nature. It involves visual receptor probability summation, individually variable decision criteria, and various other non-controllable population variables, and is defined on the basis of a threshold criterion. In psychophysics, this threshold criterion is commonly 75% correct. In a context of medical practice, as already stated here above, the desired threshold criterion would definitely have to be 100% correct. The technique introduced here is well-tailored for the pre-analysis of large bodies of medical images from patients since it allows the automatic detection of subtle but significant changes in time series of images likely to reflect growing or receding lesions. In clinical practice, finding evidence for subtle growth through visual inspection of serial imaging can be very difficult. This is especially true for scans taken at relatively short intervals (less than a year). Visual inspection often misses the slow evolution because the change may be obscured by variations in body position, slice position, or intensity profile between scans, as noted previously [7]. In some cases, the change can be too small to be noticed, which could be detrimental for the patient's treatment. Surgeons and oncologists frequently compute the change in tumor volume by comparing the measurements of consecutive scans. When the change in tumor volume is too small and hence difficult to detect between two sequential scans, neuro-radiologists tend to compare the most recent scan with the earliest available image to find any visible evidence for the evolution of the tumor. The resulting analysis does not reflect the current development of the tumor but rather a retrospective perspective of the tumor evolution [7]. This study deals with this situation and hence it can aid clinicians in deciding treatment. The *QE* is a quality measure for SOM. It is therefore expected to produce same values when the initial SOM settings and parameters remain the same and there are no changes in the input vector (image). When, the image data is altered and the SOM parameters are not altered, changes in the *QE* can reasonably be attributed to the developments taking place in the organ whose image is under study. This is why the *QE* is proposed here as a clinical determinant of the progression or remission of lesions in medical images. The process of executing the code to determine the *QEs* of twenty images takes about 40 seconds. This involves reading the DICOM images from a folder, running the SOM and determining the *QE* for each image, displaying the image on the screen and saving the *QE* value in a text file. Further studies will be performed comparing results from real patient data in the light of analyses by human experts and metrics proposed by the World Health Organization [12].

## 7. Conclusions

When the *QE* from SOM on a patient's medical images taken at different consecutive times rises, it is a potential indication that lesions or impurities on the organ under study are increasing, while a decrease may indicate the lesions are receding. To the best of our knowledge, our approach is the first to automatically detect potentially critical local changes in a patient by comparing images taken during subsequent clinical visits without relying on visual inspection or manual annotations. Our method detects changes rapidly with a minimal computation time using consecutive images of an organ without having to rely on image qualities that are derived from previous images, as is the case for image subtraction methods. This represents a clear advantage compared with monitoring cancer progression/remission via manual segmentation of several images in an MRI sequence, which is prohibitively time consuming. Automatic segmentation is a challenging, but computationally expensive task and may result in high estimation errors. Our method analyzes image contents in real-time, directly on the basis of image statistics generated through a self-organizing machine learning technique. It is demonstrated that the *QE* value of the output of these analyses "detects" the smallest increase in potentially relevant local image contents, impossible for human non-experts to detect, as testified by hit rates well below the 75% threshold limit.

**Funding**

This research did not receive any specific grant from funding agencies in the public, commercial, or not-for-profit sectors.

**Figures and Tables with legends**

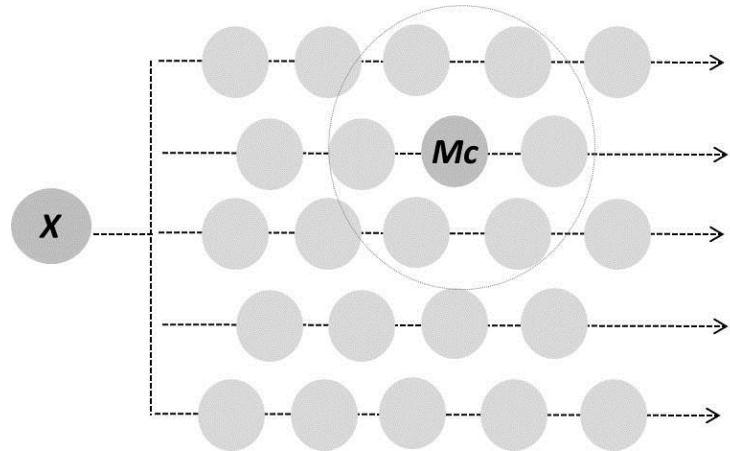

Figure 1: Schematic illustration of a self-organizing map (SOM). An input data item ($X$) is transmitted to an ensemble of models of which model $M_c$ matches best with $X$. Models that lie in the neighborhood (indicated by the large circle here) of $M_c$ in the map match better with X than all others (illustration adapted from [8]).

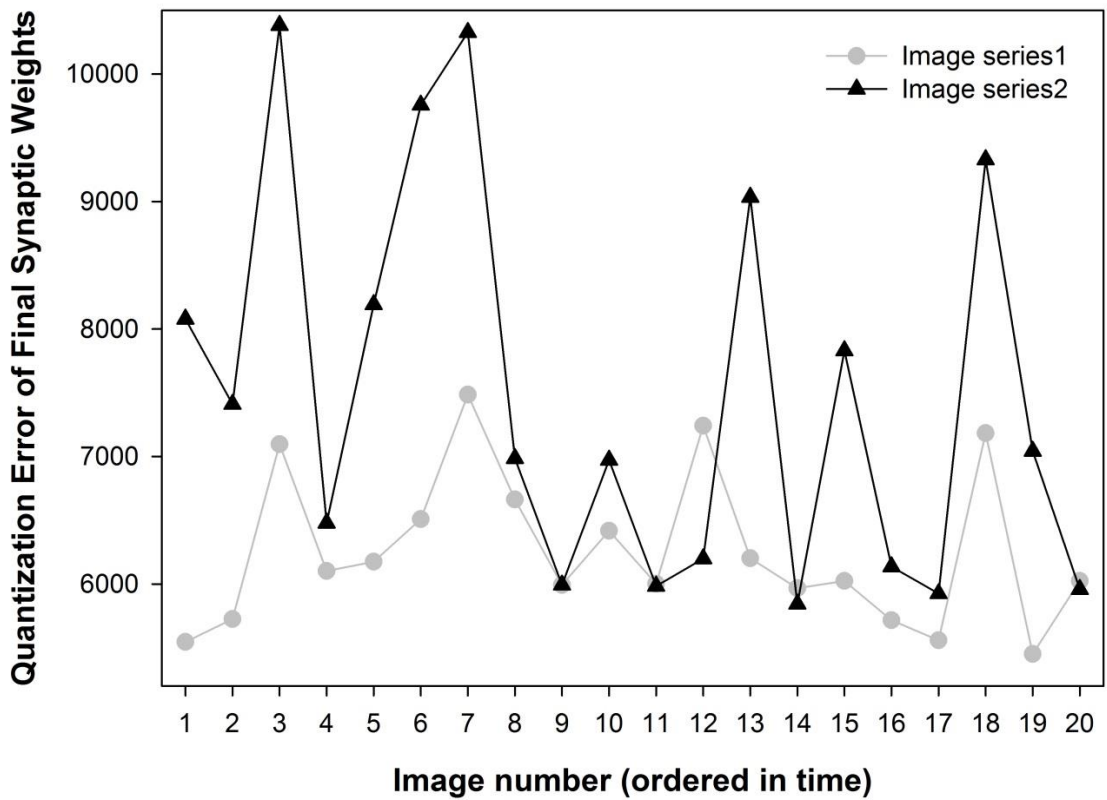

Figure 2: Results from SOM analyses on time series of the original knee images, taken at two different moments in time, before (series 1) and after (series 2) blunt force trauma. It is shown that the *QE* increases significantly (t (1, 38) = 3,336; p<.01) between image series taken before and after the clinically significant event, which signifies that the QE is a statistically highly reliable detection measure of the changes between     the images from the two series.

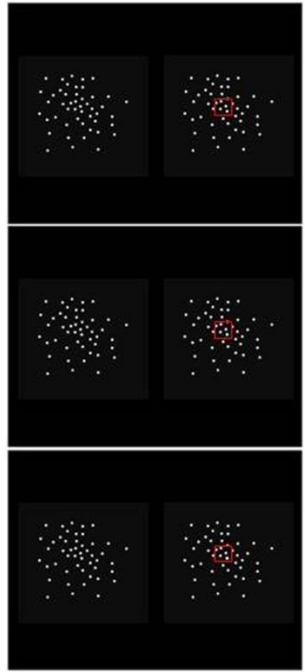

Figure 3: Random dot-images with different percentages of artificially induced and strictly local "lesion" contents (5%, 10 % and 30 % increase in the size of a single small dot, shown here highlighted by the red square) were paired with images where no such local "lesion" was added (images on left in a given pair here above). Right and left images in a pair varied between presentations, in random order. Equivalent proportions of pairs with two identical images (not shown here) were also presented in the task sequence to measure the tendency of an individual to give false alerts ("guess rates") by responding "different" when they should respond "same".

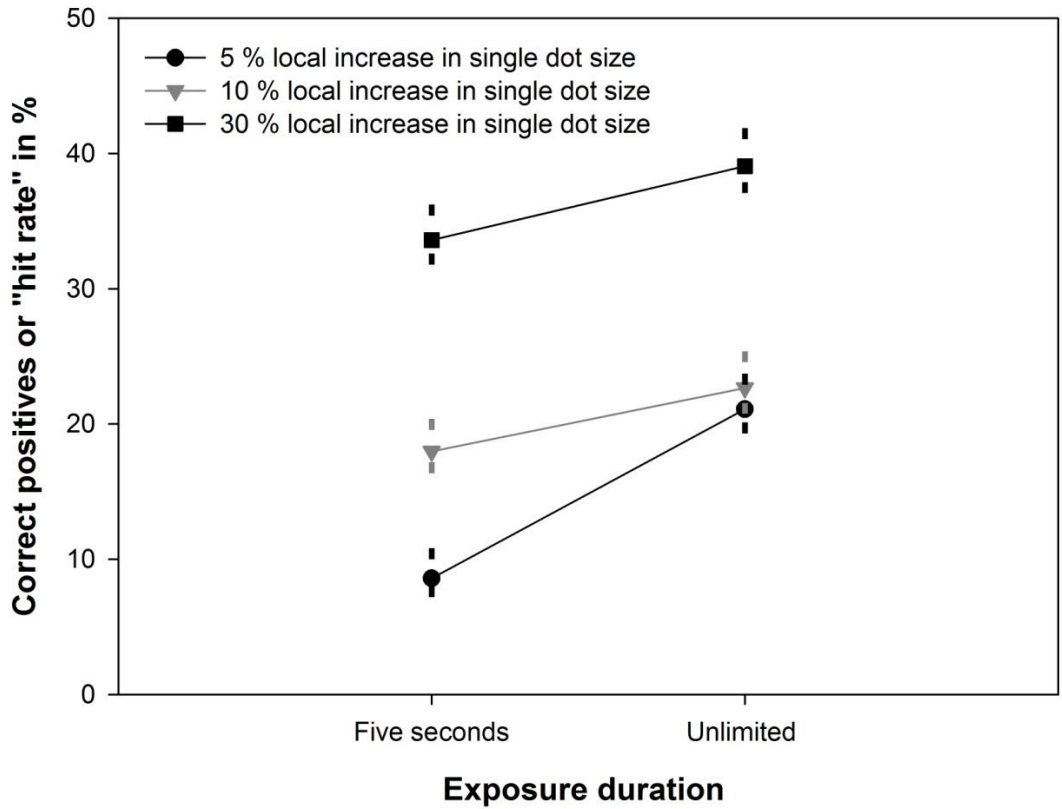

Figure 4: Average correct positive (*CP)* rates and standard errors as a function of the % increase in local "lesion" content and the exposure duration of image pairs during the experiment.

*Image pairs with five seconds exposure*

|   |   | SAME | DIFFERENT |
|---|---|---|---|
|   | "same" | **89** *(CN)* | **91** *(FN)* |
| **R** | "different" | **11** *(FP)* | **9** *(CP)* |

a)

*Image pairs with observer controlled ("unlimited") exposure*

|   |   | SAME | DIFFERENT |
|---|---|---|---|
|   | "same" | **86** *(CN)* | **91** *(FN)* |
| **R** | "different" | **14** *(FP)* | **9** *(CP)* |

b)

Table 1: Conditional response rates (in %) of the novices for "no change" (same) image pairs and image pairs with a 5% local single dot size increase in one of the pair under conditions of five seconds exposure duration *(a)*, and observer controlled exposure duration *(b)* for each image pair. Correct positive rates (*CP*), often also called "hits rates", correct negative (*CN*), false positive (*FP*), and false negative (*FN*) rates are shown.

*Image pairs with five seconds exposure*

|   |           | SAME      | DIFFERENT |
|---|-----------|-----------|-----------|
|   | "same"    | **88** *(CN)* | **82** *(FN)* |
| ***R*** | "different" | **12** *(FP)* | **18** *(CP)* |

a)

*Image pairs with observer controlled ("unlimited") exposure*

|   |           | SAME      | DIFFERENT |
|---|-----------|-----------|-----------|
|   | "same"    | **87** *(CN)* | **77** *(FN)* |
| ***R*** | "different" | **13** *(FP)* | **23** *(CP)* |

b)

Table 2: Conditional response rates (in %) of the novices for "no-change" (same) image pairs and pairs where one of the images contained a 10% local size increase of a single dot, under conditions of five seconds exposure duration *(a)*, and observer controlled exposure duration *(b)*.

*Image pairs with five seconds exposure*

|   |           | SAME      | DIFFERENT |
|---|-----------|-----------|-----------|
|   | "same"    | **86** *(CN)* | **66** *(FN)* |
| **R** | "different" | **15** *(FP)* | **34** *(CP)* |

a)

*Image pairs with observer controlled ("unlimited") exposure*

|   |           | SAME      | DIFFERENT |
|---|-----------|-----------|-----------|
|   | "same"    | **87** *(CN)* | **61** *(FN)* |
| **R** | "different" | **13** *(FP)* | **39** *(CP)* |

b)

Table 3: Conditional response rates (in %) of the novices for "no-change" (same) image pairs and pairs with a 30% local dot size increase in one of the images under conditions of five seconds exposure duration *(a)*, and observer controlled exposure duration *(b)*.

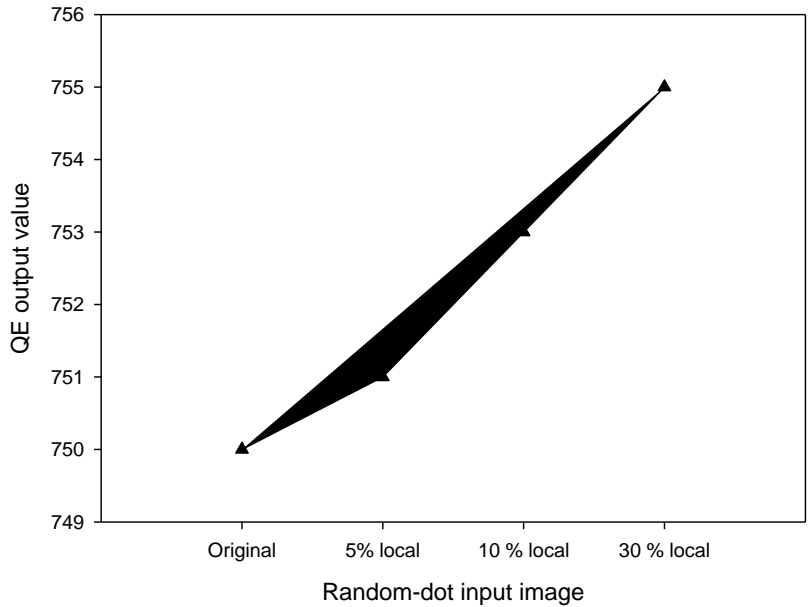

Figure 5: *QE* values from constant parameter SOM on the random-dot images as a function of image contents. Computer output for the original (no change) image and the images with 5%, 10%, and 30% increase in single local dot size shows. The *QE* increases systematically and reliably (no variation in *QE* across multiple constant parameter SOM for one and the same image) with the increase in single local dot size in the input image.

*Image pairs with five seconds exposure*

|   |           | SAME         | DIFFERENT    |
|---|-----------|--------------|--------------|
|   | "same"    | **41.7** *(CN)* | **58.3** *(FN)* |
| **R** | "different" | **58.3** *(FP)* | **41.7** *(CP)* |

a)

*Image pairs with observer controlled exposure*

|   |           | SAME         | DIFFERENT    |
|---|-----------|--------------|--------------|
|   | "same"    | **42** *(CN)*  | **41** *(FN)*  |
| **R** | "different" | **58** *(FP)*  | **59***(CP)*   |

b)

<u>Table 5</u>. The conditional detection rates in % of the experts for random-dot images with a 5% increase in local size of a single dot in the image. There is no detection of this change by the experts, they are basically guessing. The CP rate after subtraction of the FP rate is almost zero. Note that the guess rate (FP rate) of the three experts here is noticeably higher than was that of the novices. Expert radiologists thus seem to have a stronger tendency to produce false positives ("better safe than sorry" strategy).

*Image pairs with five seconds exposure*

|  |  | SAME | DIFFERENT |
|---|---|---|---|
|  | "same" | **42** *(CN)* | **50** *(FN)* |
| *R* | "different" | **58** *(FP)* | **50** *(CP)* |

a)

*Image pairs with observer controlled exposure*

|  |  | SAME | DIFFERENT |
|---|---|---|---|
|  | "same" | **41** *(CN)* | **17** *(FN)* |
| *R* | "different" | **59** *(FP)* | **84** *(CP)* |

b)

<u>Table 6</u>. Conditional detection rates in % of the experts for random-dot images with a 10% increase in local dot size. The experts are still basically guessing, but beginning to detect the difference in the two images when they can control the exposure duration: CP rate – FP rate = 25.

*Image pairs with five seconds exposure*

|   |   | SAME | DIFFERENT |
|---|---|---|---|
| **R** | "same" | **42** *(CN)* | **16** *(FN)* |
|   | "different" | **58** *(FP)* | **84** *(CP)* |

a)

*Image pairs with observer controlled exposure*

|   |   | SAME | DIFFERENT |
|---|---|---|---|
| **R** | "same" | **41** *(CN)* | **8** *(FN)* |
|   | "different" | **59** *(FP)* | **92** *(CP)* |

b)

<u>Table 7</u>. Conditional detection rates in % of the experts for random-dot images with a 30% local dot size increase. The experts, like the novices, better detect a 30% difference in local dot size between the two images when they can control exposure duration. CP rate – FP rate = 33.3.